\documentclass[letterpaper]{article} 
\usepackage{aaai2026}  
\usepackage{times}  
\usepackage{helvet}  
\usepackage{courier}  
\usepackage[hyphens]{url}  
\usepackage{graphicx} 
\usepackage{natbib}  
\usepackage{caption} 
\usepackage{algorithm}
\usepackage{algorithmic}
\usepackage{booktabs}

\usepackage{newfloat}
\usepackage{listings}
\DeclareCaptionStyle{ruled}{labelfont=normalfont,labelsep=colon,strut=off} 
\floatstyle{ruled}
\newfloat{listing}{tb}{lst}{}
\floatname{listing}{Listing}
\title{Beyond Blind Spots: Analytic Hints for Mitigating LLM-Based Evaluation Pitfalls}
\author{Ora Nova Fandina, Eitan Farchi, Shmulik Froimovich, Raviv Gal \\
Wesam Ibraheem, Rami Katan, Alice Podolsky \\}
\affiliations{IBM Research, Israel}

\usepackage{bibentry}

\begin{document}

\maketitle

\begin{abstract}
Large Language Models are increasingly deployed as judges (LaaJ) in code generation pipelines. While attractive for scalability, LaaJs tend to overlook domain-specific issues raising concerns about their reliability in critical evaluation tasks. To better understand these limitations in practice, we examine LaaJ behavior in a concrete industrial use case: legacy code modernization via COBOL code generation. In this setting, we find that even production-deployed LaaJs can miss domain-critical errors, revealing consistent blind spots in their evaluation capabilities.

To better understand these blind spots, we analyze generated COBOL programs and associated LaaJs judgments, drawing on expert knowledge to construct a preliminary taxonomy. Based on this taxonomy, we develop a lightweight analytic checker tool that flags over 30 domain-specific issues observed in practice. We use its outputs as {\it analytic hints}, dynamically injecting them into the judge’s prompt to encourage LaaJ to revisit aspects it may have overlooked.

Experiments on a test set of 100 programs using four production-level LaaJs show that LaaJ alone detects only about 45-63\% of the errors present in the code (in all judges we tested), while the analytic checker alone lacks explanatory depth. When combined, the LaaJ+Hints configuration achieves up to 74\% coverage (for the best-performing judge and injection prompt) and produces qualitatively richer, more accurate explanations, demonstrating that analytic–LLM hybrids can substantially enhance evaluation reliability in deployed pipelines. We release the dataset and all used prompts. 
\end{abstract}

\section{Introduction}
As code generation systems improve, evaluation must keep pace, not just in scale but in depth. The ability to assess the correctness, safety, and relevance of generated code is critical for real-world deployment, especially in high-stakes domains. Large Language Models (LLMs), when used as LLM-as-a-Judge offer a scalable alternative to human or automatic analytic evaluation, particularly for tasks lacking clear ground truth. While LLMs have demonstrated strong general reasoning capabilities needed for judgment , prior studies suggest they often struggle with tasks requiring deep domain knowledge. Our work provides further empirical support for this observation, focusing on how reliably LaaJs perform in real-world, domain-specific evaluation scenarios.

We focus on COBOL code generation, a representative task in legacy system modernization where domain-specific evaluation is especially challenging. Unlike modern languages, COBOL often involves non-standard control flow, implicit data handling, and business-specific conventions that are rarely documented or standardized. These patterns are difficult for general-purpose LLMs to recognize, especially given COBOL’s under-representation in training data and the relatively sparse availability of public benchmarks or test suites.

\begin{figure}[t]
  \centering
  \includegraphics[width=\linewidth]{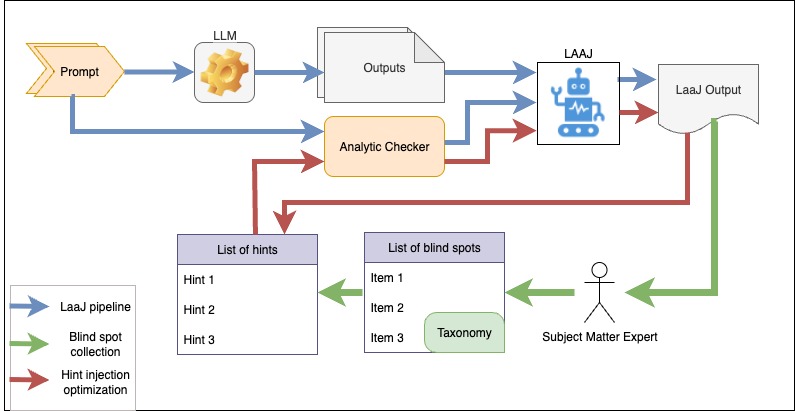}
  \caption{Full Workflow for improving LLM-as-a-Judge evaluations. The main pipeline (blue) generates candidate outputs from the LLM and scores them using a LaaJ. Two refinement loops are overlaid: (1) {\bf Blind Spots Collection} (green), where SMEs analyze LaaJ errors and curate a taxonomy of blind spots, and (2) {\bf Hints Optimization} (red), where targeted hints are derived and injected via an Analytic Checker to guide LaaJ reasoning. This process allows LaaJ evaluations to be continuously improved by addressing specific weaknesses in model judgment.}

  \label{fig:wokflow}
\end{figure}

In our internal setting, we developed and deployed multiple LaaJ configurations specifically tailored for COBOL evaluation task. Although these LaaJs were carefully engineered and refined through human-in-the-loop testing, we still found that critical domain-specific issues were often missed.
We hypothesized that these failures stem not merely from insufficient tuning, but from {\bf blind spots}, recurrent omissions or reasoning failures that current LaaJs consistently overlook. These blind spots are often subtle: a missing initialization, a file operation without proper status checks, or improper use of restart logic. While each may appear minor in isolation, collectively they undermine the reliability of model-based evaluation in high-stakes environments.

To systematically investigate these limitations, we designed a multi-stage workflow (Figure 1) involving LaaJ error analysis, blind spot identification, and targeted intervention. We first constructed a curated dataset of COBOL programs containing realistic, domain-specific issues, and used production-grade LaaJs to evaluate them. Expert reviewers then analyzed LaaJ judgments to identify failure modes, resulting in a structured taxonomy of blind spots. Based on this taxonomy, we developed an analytic checker that uses a catalog of over 30 issue types to dynamically analyze each generated program. For each input, it emits concise, structured hints for the detected issues, which are then injected into the LaaJ prompt to guide the model toward known blind spots.
This approach offers a practical, non-intrusive mechanism for improving evaluation depth: by dynamically guiding the model’s focus for each input at inference time, it helps mitigate per-instance evaluation failures without retraining or fine-tuning, while preserving auditability and compatibility with existing production pipelines.

Our main contributions are: 
\begin{itemize}

\item Expert-guided discovery and categorization of LaaJ blind spots in COBOL evaluation; the initial taxonomy;

\item A rule-based analytic checker that scans the code and outputs analytic hints;

\item A demonstration that prompt-level analytic hint injection yields measurable improvements in LaaJ performance.

\end{itemize}

We release the dataset, prompts and analytic checker. \footnote{\url{https://github.com/AlicePod1/analytic-hints-paper}}

\paragraph{Related Work}
LLMs are increasingly used as automated judges in code generation \cite{zhuo-2024-ice,tong-zhang-2024-codejudge, Zhao2024CodeJudgeEvalCL}. 
However, recent studies have shown that LaaJs often overlook fine-grained errors and struggle to capture semantic correctness, particularly in complex domains like code generation \cite{doddapaneni2024-fbi,tong-zhang-2024-codejudge,tan2024-judgebench}. 
To address these issues, several works proposed enhancing LaaJ reasoning with prompt-level interventions, such as self-refinement \cite{MadaanSelfRefine}, rationale rewriting \cite{trivedi2024self}, or tool-assisted evaluation \cite{toolformer}. 

Closest to our approach are hint-injection methods that steer LLM reasoning with targeted cues, including AutoHint (learned task-specific hints), Directional Stimulus Prompting (policy-generated stimulus tokens) \cite{stimulusPrompting}, Progressive-Hint Prompting (reusing prior answers as hints) \cite{zheng2023progressive}, and Hint-before-Solving (pre-drafted hints) \cite{Fu2024HintbeforeSolvingPG}. 

Unlike these approaches (mostly evaluated on general reasoning/math), our method injects expert-validated, rule-based analytic hints extracted directly from the code, yielding interpretable, high-precision guidance for LaaJ without any auxiliary policy model, self-hinting loop, or retraining.

\section{From Failure Analysis to Hint-Guided Evaluation} 
\noindent 
\paragraph{Identifying Blind Spots}  We constructed a development dataset of 100 COBOL programs deliberately generated to include realistic, domain-specific errors. The programs were produced using the \texttt{mistral-medium-2505} language model with a carefully crafted instruction prompt. Human experts subsequently validated that the generated code indeed exhibited the intended faults.

We evaluated the generated COBOL programs using two LaaJ configurations that were developed and deployed as part of our production pipeline for automated code quality assessment. These configurations are based on the following LLMs:\texttt{llama-3-405b}, and \texttt{mistral-medium-2505}.

Drawing on domain expertise, we manually analyzed the outputs to identify and characterize recurring blind spots of the judges. This analysis was conducted in iterative refinement cycles: experts reviewed missed issues, updated annotations, and adjusted emerging categories as patterns became clearer. Over multiple passes, a stable set of recurring failure types emerged. We distilled these into a structured taxonomy capturing the most characteristic classes of LaaJ evaluation failure.
\noindent
\paragraph{Taxonomy} Below we present an initial taxonomy comprising six categories. Each reflects a class of domain-specific evaluation challenges that were frequently missed by LaaJs, even after the rigorous prompt tuning and human-in-the-loop validation we conducted while developing these production-level judges.
\begin{figure}[t]
  \centering
  \includegraphics[width=\linewidth]{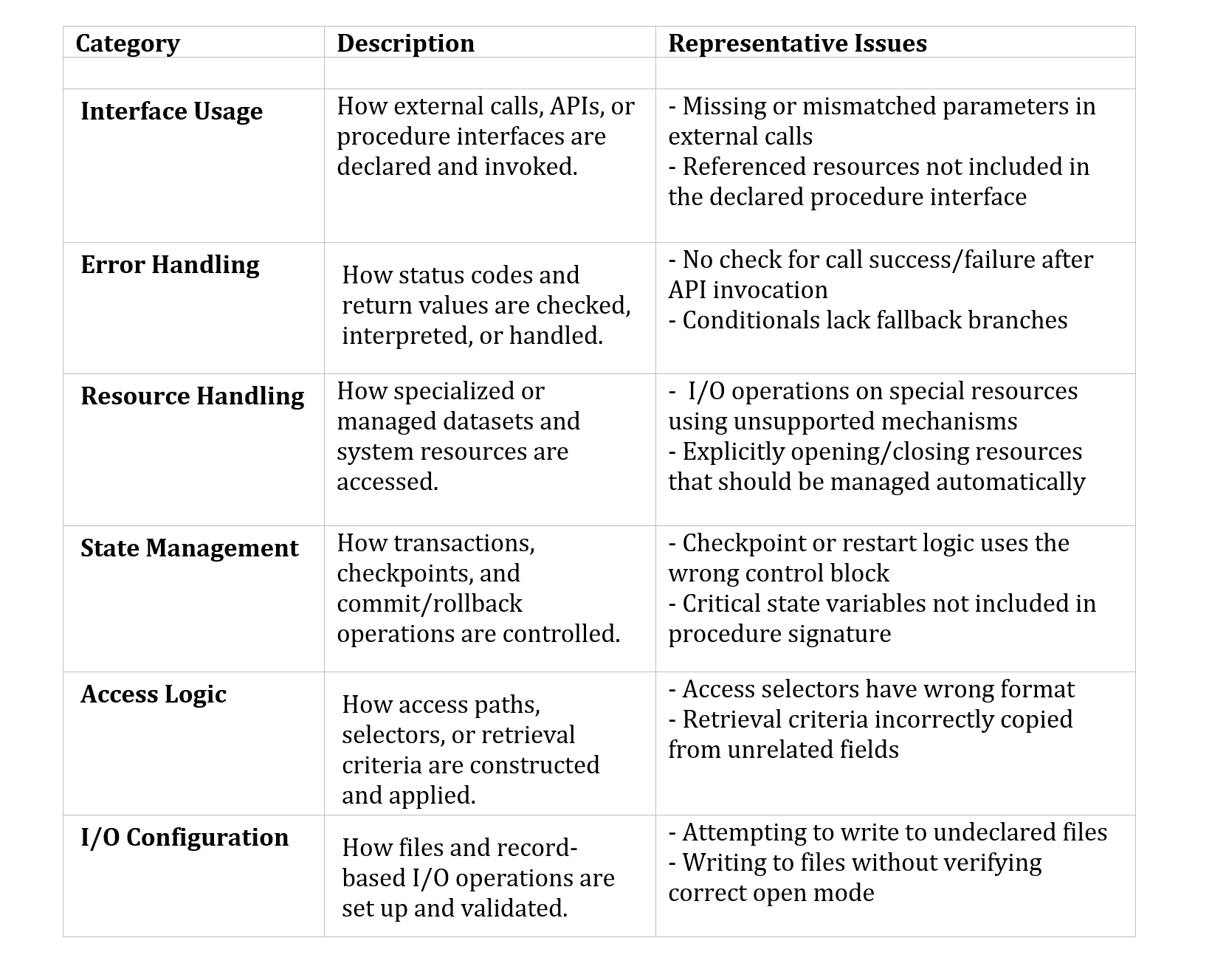}
  \caption{Taxonomy of domain-specific evaluation issues identified in COBOL LaaJ failures.}
  \label{fig:taxonomy}
\end{figure}
 These categories reflect patterns that emerged repeatedly across expert reviews of LaaJs evaluations. While some relate to syntactic or structural violations (e.g., missing status fields, undeclared descriptors), others reveal deeper semantic blind spots that require non-local reasoning about state, control flow, or implicit conventions in COBOL-IMS (Information Management System) systems.

These categories, while rooted in COBOL-specific evaluation and the particular set of base LLM models, reveal broader patterns that may generalize to other domains and LaaJs. Across categories, we observe that LaaJs tend to struggle with {\bf multi-line reasoning}, particularly when issues depend on non-local context or the interaction between distributed control structures. Several error types reflect {\bf omissions} rather than incorrect content, such as missing status checks, or initializations, which highlights LaaJ's difficulty in detecting when something important is not present. This points to a broader limitation:  foundation models often excel at recognizing what is in the input, but struggle to reason about what should be there and is missing. Finally, many errors stem from misunderstanding {\bf execution order} or control flow, such as using a file before it is opened or skipping necessary status checks after a call. These cases suggest that current LaaJs lack the ability to simulate program semantics or execution state, instead relying heavily on surface-level patterns.
Taken together, these observations may offer insights into the limitations of current LLM-based code quality evaluators.

\noindent
\paragraph{Analytic Checker and Hints}  We developed a lightweight analytic checker that encodes over 30 error types identified through expert review. The checker uses pattern-matching to detect the issues in the program and emits short, human-readable messages, we call analytic hints. These hints are then dynamically injected into the LaaJ prompt to help overcome its blind spots, encouraging the model to revisit aspects of the input code it previously overlooked. While our injection strategy is deliberately naive, plain-text hints placed at the top of the judge's prompt, it already leads to measurable improvements in LaaJ performance. This opens a path for further refinement: by tuning the phrasing, formatting, or placement of the hints, developers may unlock additional gains with minimal overhead. 



\section{Experiments and Results}

\paragraph{Setup}
We constructed a test set of 100 fresh synthetic COBOL programs, each deliberately seeded with multiple subtle errors. This dataset was reserved exclusively for evaluation and was not used during the taxonomy construction or tool development phases. The programs were generated with \texttt{mistral-medium-2505} using a carefully crafted instruction prompt, and human experts validated that the intended errors are present.

We evaluated four production LaaJ configurations, based on \emph{llama-4-Maverick}, \emph{llama-3-405b}, \emph{DeepSeek-v3}, and \emph{gpt-oss-120b}, from our internal pipeline for automated code quality assessment. All judges use the same detailed evaluation prompt (not disclosed due to proprietary constraints), which we refer to as the {\it native prompt} in our experiments. 

The hint injection was implemented in two ways. First, in a naïve setup, per-input hints were inserted by simply appending them to the end of the native prompt. Second, in a more guided setup, the prompt explicitly instructed the model to address the provided hints, in a detailed way. We used these two configurations to demonstrate how prompt design influences the effectiveness of the hint-injection phase.

\paragraph{Results}
For each judge, we conducted the following experiment. For every sample COBOL program, we first estimated the total number of issues present. Since the exact number of true issues is unknown, we approximated it as the union of errors detected by the analytic tool and by the native judge, that is the sum of both counts minus their overlap. We then run the hybrid judges on each sample: naive hint injection prompt and best hint injection prompt.

 \begin{figure}
    \centering
    \includegraphics[width=1\linewidth]{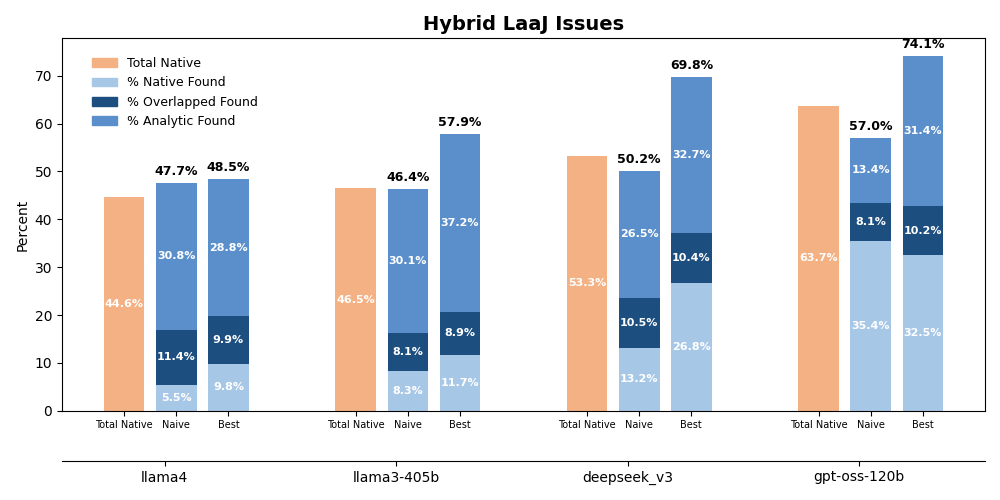}

    \caption{Hybrid Laajs detection rates (naive vs optimized hint injection schemes), with Native Laajs detection rates.}
    \label{fig:main_exp}
\end{figure}

We then evaluated the hybrid judges on each sample using two configurations: the naïve hint-injection prompt and the optimized (best) hint-injection prompt. For each configuration, we measured the fraction of errors detected by the hybrid judge within each error category: analytic errors, overlapping errors, and errors originally detected by the native judge, and averaged the results over all 100 programs to obtain the final coverage rates.

As shown in Figure \ref{fig:main_exp}, introducing analytic hints led to consistent gains across all models. Native judges detected roughly 45–63\% of total issues, whereas hybrid judges reached 48–74.1\%, depending on model and prompt. Most of the improvement stems from capturing analytic-only errors previously missed by the native judges, while performance on native-only and overlap categories remained stable or improved slightly. We observe that the optimized hint injection achieves consistently high success rates of issues detected with analytic checker and confirmed by LaaJ, with DeepSeek-v3 and gpt-oss-120b exceeding 90\%.



\begin{figure}
    \centering
    \includegraphics[width=1\linewidth]{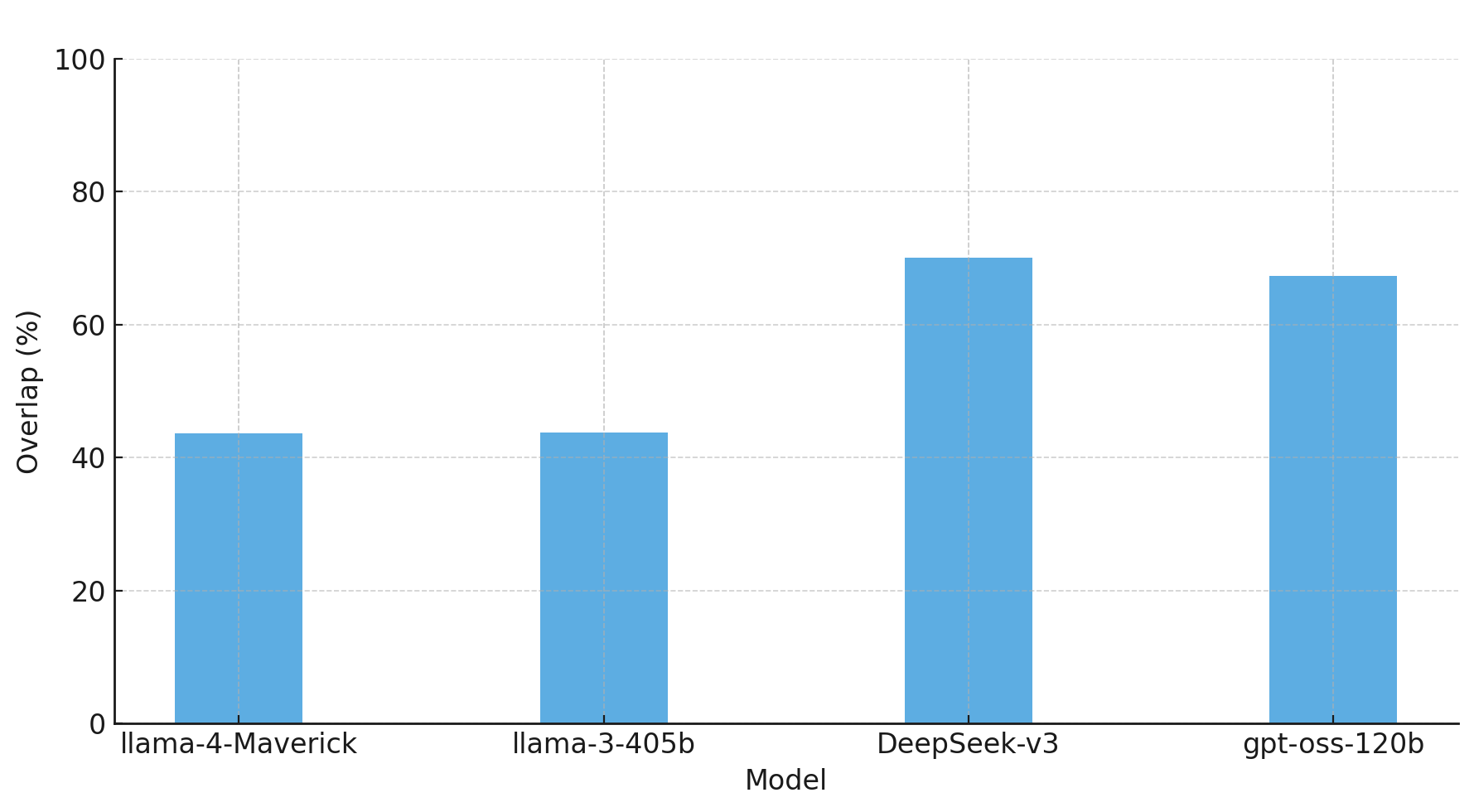}
    \caption{Percent of native judge errors rediscovered by hybrid judges (with best hint-injection prompt).}
    \label{fig:retained_errors}
\end{figure}

To verify that the hint-augmented judges retained their original evaluative capabilities (specifically, the ability to rediscover issues not explicitly listed in the analytic hints) we performed an additional analysis.
We converted the explanations produced by the native judges into structured lists of issues and then searched for corresponding or semantically similar issues within the explanations of the hint-augmented judges.
This matching was conducted by a \textit{gpt-oss-120b}-based issue finder judge. The results, summarized in Figure \ref{fig:retained_errors}, show that the hint-augmented judges reproduced approximately 45–70\% of the issues originally identified by their native counterparts. In particular, DeepSeek-v3 and gpt-oss-120b achieved the highest rediscovery rates, recovering 70.06\% and 67.31\% of the native-judge issues, respectively. 

Interestingly, the rediscovery rates varied considerably across models, indicating that some judges preserved their original evaluative behavior more effectively than others when augmented with analytic hints. We propose using this issue-rediscovery rate as an additional diagnostic metric for assessing evaluator reliability: higher values indicate models that can integrate new analytic guidance without losing prior evaluative competence, whereas lower values may reveal instability or over-dependence on prompt conditioning.

\section{Conclusions}
We presented a practical approach for enhancing LaaJs using analytic hint injection in code evaluation. Grounded in expert analysis of COBOL evaluation failures, our method couples a taxonomy of blind spots with a lightweight checker that emits targeted hints. Injecting these hints into the judge’s prompt refocuses its reasoning toward previously overlooked issues, yielding significant gains without retraining.
Among all models, the \textbf{gpt-oss-120b} judge with hints achieved the best performance, addressing \textbf{74.1\%} of errors while retaining \textbf{67.31\%} of those identified by the native judge.
This demonstrates that prompt-level analytic interventions can substantially improve judgment coverage while preserving general evaluative capabilities. 

 Notably, the hint-augmented judges did not fully reproduce all issues identified by their native counterparts, with rediscovery rates of 45\% and 67\%. This drop suggests that the injected hints altered the judges’ focus: by emphasizing analytic issues, they improved detection of these issues but deprioritized unhinted aspects of the evaluation. Thus, hint injection improves targeted diagnostic precision but may narrow overall coverage, highlighting both the promise and the limitation of analytic-guided evaluation: it can direct the model's attention where it matters most but my narrow its coverage if not carefully balanced.

Our study, limited to one task and a small dataset, offers an initial demonstration that analytic guidance at inference time can substantially enhance model-based evaluation.
This work contributes to the growing body of research on \textit{hybrid evaluation systems} that augment foundation models with structured analytic tools.


\bibliography{aaai2026}
\end{document}